# Influence of chirp, jitter and relaxation oscillations on probabilistic properties of laser pulse interference

Roman Shakhovoy, Violetta Sharoglazova, Alexander Udaltsov, Alexander Duplinskiy, Vladimir Kurochkin, and Yury Kurochkin

*Abstract*— Interference of laser pulses is an essential ingredient of quantum randomness; therefore, probabilistic properties of laser pulses gains new relevance. Here, we consider in detail the combined influence of the three effects – chirp, jitter and relaxation oscillations – on the probability density function of the interference of pulses from a gain-switched semiconductor laser. We develop a rigorous model based on laser rate equations and demonstrate that only consideration of all these three effects together allows describing the interference statistics properly. We supplement our theoretical calculations with corresponding measurements at various pump currents. Experimental results demonstrate perfect agreement with predictions of the model and are well reproduced by Monte-Carlo simulations.

*Index Terms*—Interference, chirp, jitter, optical pulses, semiconductor lasers.

## I. Introduction

LASER pulse interference is an essential ingredient of quantum technologies. Weak coherent states (attenuated laser pulses) are widely used to mimic interference between single photons, which is generally employed in such quantum information applications as quantum teleportation [1], linear optics computing [2] and detector-safe quantum cryptography [3], [4]. Interference of intense coherent states, in turn, is often used in optical quantum random number generators (QRNGs) [5]-[9], where phase randomness between pulses of a gain-switched semiconductor laser acts as a source of quantum entropy. In both "single-" and multi-photon cases, the interference of laser pulses often has a number of unpleasant features, which adversely affect the visibility and have an impact on the appearance of the probability density function (PDF) of the random interference signal.

Detailed understanding of physical processes underlying the operation of an optical QRNG is vital in terms of its security. Thus, quantum noise extraction from the interference of laser pulses, which we discuss in detail in [9], requires to know the signal PDF together with its origin. Simple physical considerations usually used for interpretation of the real PDF [6], [7] could at least be confusing or may even lead to an incorrect estimate of the quantum noise contribution. So, one needs an appropriate model, which could explain the influence of laser pulse imperfections on the signal PDF.

In addition, QRNG application in quantum key distribution (QKD) requires high random bit generation rate for state preparation, which, in turn, imposes high demands on the rate of laser pulse generation. At high modulation frequencies (few GHz) of a gain-switched semiconductor laser, one has to work with short laser pulses, i.e. with the part of an optical signal at the onset of lasing, which is most affected by chirp and relaxation oscillations. Together with jitter, these effects demonstrate significant contribution to the interference and must be thus taken into account.

The combined influence of chirp and jitter in the context of QRNG was considered in [7], where authors demonstrated that the PDF of the interference signal for chirped laser pulses differs markedly from the PDF measured in the absence of chirp. The authors proposed a simple model, in which laser pulses were assumed to have a Gaussian shape and exhibit a linear chirp. The contribution of relaxation oscillations was not taken into account in their considerations, moreover, authors made an assumption of a uniform PDF for the jitter needed to fit theoretical results to experimental data. However, distribution of pulse emission time fluctuations can be shown to be quite close to Gaussian in case of gain-switched lasers [10] and usually has the rms of the order of 10-50 ps [11], [12]. More detailed investigation reveals that the appearance of the interference signal PDF with more realistic jitter is different from the one shown in [7] and one should take into account relaxation oscillations in order to agree theory and experiment.

This work was supported by Russian Science Foundation (Grant No. 17-71-20146). *(Corresponding author: Roman Shakhovoy.)*

R. Shakhovoy, A. Udaltsov, A. Duplinskiy, and V. Kurochkin are with Russian Quantum Center, 121205 Moscow, Russia, and also with QRate, 121353 Moscow, Russia (e-mail: r.shakhovoy@goqrate.com).

V. Sharoglazova is with Russian Quantum Center, 121205 Moscow, Russia, QRate, 121353 Moscow, Russia, and also with Skolkovo Institute of Science and Technology, 121205 Moscow, Russia.

Y. Kurochkin is with Russian Quantum Center, 121205 Moscow, Russia, QRate, 121353 Moscow, Russia, and also with NTI Center for Quantum Communications, National University of Science and Technology MISiS, 119049 Moscow, Russia.

In the present work, we develop a rigorous model based on laser rate equations, which considers interference of laser pulses in the presence of relaxation oscillations, jitter and chirp. We show that only inclusion of all these three effects into consideration allows explaining evolution of the signal PDF against the pump current. To the best of our knowledge, such a discussion of laser pulse interference in the context of its probabilistic properties has not been previously covered in the literature. To demonstrate the efficiency of the proposed model, we supplement our theoretical calculations with corresponding measurements. Experimental results are in perfect agreement with calculations and are well reproduced by Monte-Carlo simulations.

## II. THEORY

Let us first consider the laser pulse interference measured using a Michelson fiber optic interferometer (Fig. 1). The delay line $\Delta L$ is chosen so that the corresponding delay time defined by $\Delta T = 2\Delta L n_g / c$ is multiple of the pulse repetition period $2\pi/\omega_p$, such that at the output of the interferometer the $i$-th laser pulse of the sequence meets the $i+N_p$-th pulse, where $N_p$ is the number of pulses that pass the short arm during the time needed for the pulse to pass the long arm (here $c$ is the light speed in vacuum and $n_g$ is the group index).

In the following, subscripts 1 and 2 will be used to designate the short and long arms of the interferometer, respectively, and to designate laser pulses coming from the respective arms. Assuming that interfering pulses are polarized in the same plane, the intensity of the signal at the output of the interferometer can be written as follows:

$$S(t) \sim \left| E_1(t) + E_2(t) \right|^2, \quad (1)$$

where $E_1$ and $E_2$ are (scalar) electric fields in the first and second pulses, respectively. The time dependence of the electric field in a pulse can be written in the following form:

$$E_{1,2}(t) \sim \sqrt{P_{1,2}(t)} e^{i\varphi_{1,2}(t)}, \quad (2)$$

where $\varphi_{1,2}(t)$ is the phase of the field and $P_{1,2}(t)$ is the output power in the corresponding pulse. The power of the interfering pulses can be related to the laser output power $P(t)$ as follows:

$$\begin{aligned} P_1(t) &= (1-a_1)T_{01}T_{10}P(t), \\ P_2(t) &= (1-a_2)T_{02}T_{20}P(t-\Delta t), \end{aligned} \quad (3)$$

where $a_1$ and $a_2$ stand for the losses in the optical fiber in the short and long arms, respectively, $T_{kl}$ is a coupler transmittance from the input port $k$ to the output port $l$ (see Fig. 1), and where we introduced the inaccuracy of the pulse overlap $\Delta t$, i.e. we took into account that one of the pulses may exit the interferometer a bit earlier than the other.

Time evolution of the power $P$ and the phase $\varphi$ of the electric field in the laser pulse can be found from the system of standard laser rate equations [13]-[16]:

$$\begin{aligned} \frac{dQ}{dt} &= (G-1)\frac{Q}{\tau_{ph}} + C_{sp}\frac{N}{\tau_e}, \\ \frac{d\varphi}{dt} &= \frac{\alpha}{2\tau_{ph}}(G_L - 1), \\ \frac{dN}{dt} &= \frac{I}{e} - \frac{N}{\tau_e} - \frac{QG}{\Gamma \tau_{ph}}. \end{aligned} \quad (4)$$

Here $Q$ is the absolute square of the normalized electric field amplitude corresponding to the average photon number inside the laser cavity and related to the output power by $P = Q(\varepsilon \hbar \omega_0 / 2\Gamma \tau_{ph})$, where $\hbar \omega_0$ is the photon energy ($\omega_0$ is the carrier frequency), $\varepsilon$ is the differential quantum output, $\Gamma$ is the confinement factor, $\tau_{ph}$ is the photon lifetime inside the cavity, and the factor $1/2$ takes into account the fact that the output power is generally measured from only one facet. Onwards, $N$ is the carrier number, $I$ is the pump current, $e$ is the absolute value of the electron charge, $\tau_e$ is the effective lifetime of the electron, the factor $C_{sp}$ corresponds to the fraction of spontaneously emitted photons that end up in the active mode, $\alpha$ is the linewidth enhancement factor (Henry factor [17]), and the dimensionless linear gain $G_L$ is defined by $G_L = (N-N_0)/(N_{th}-N_0)$, where $N_0$ and $N_{th}$ are the carrier numbers at transparency and threshold, respectively. The gain saturation [15] is included in Eq. (4) by using the relation $G = G_L(1-\chi P)$, where $\chi$ is the gain compression factor [13]. (Note that in Eq. (4) equations for $Q$ and $N$ contain $G$, whereas equation for $\varphi$ contains the linear gain $G_L$.)

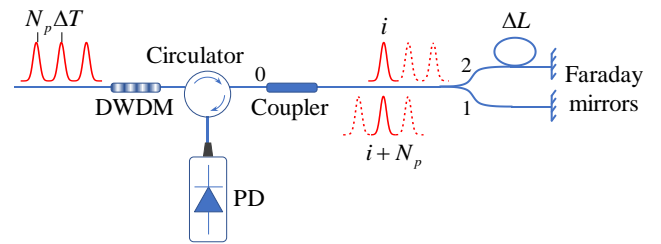

Fig 1. The optical scheme used in this work to observe interference of laser pulses. The circulator is used to separate optical signals that travel in opposite directions and thus to prevent unwanted feedback into a laser. PD stands for the photodetector; DWDM – dense wavelength division multiplexing bandpass filter. $\Delta T$ and $\Delta L$ are defined in the text.

The signal corresponding to a pair of interfering pulses can be now written in the following form:

$$S(t) \sim \left| \sqrt{p_1 P_1(t)} \exp\left[i\varphi(t) + i\varphi_{p1} + i\theta_1\right] \right. \\ \left. + \sqrt{p_2 P_2(t-\Delta t)} \exp\left[i\varphi(t-\Delta t) + i\varphi_{p2} + i\theta_2\right] \right|^2, \quad (5)$$

where $P_1$ and $P_2$ are given by Eq. (3) and the phase $\varphi(t)$ should be taken from the solution of Eq. (4). (The role of parameters $p_1$ and $p_2$ will be explained below.) It is taken into account in Eq. (5) that laser pulses acquire different phases when passing along different arms of the interferometer and the corresponding phase difference is $\Delta\theta = \theta_2 - \theta_1 = 2\Delta L n \omega_0 / c$, where $n$ is the fiber refractive index and the factor 2 stands because the pulses pass the delay line twice in the Michelson interferometer. It should be noted that the second term in Eq. (5) does not contain the factor $\exp(i\omega_0 \Delta t)$ that reflects the fact that the phase difference of the pulses does not depend on the accuracy of their overlap $\Delta t$, but is determined by the difference $\varphi_{p2} - \varphi_{p1}$ and by the interferometer delay line. Here the phases $\varphi_{p1}$ and $\varphi_{p2}$ are acquired by laser pulses during the time, when the gain switched laser is under threshold (in the amplified spontaneous emission (ASE) mode). We will assume below that phase correlations of the electromagnetic field are destroyed very quickly in the ASE mode due to contribution of phase uncorrelated spontaneous transitions, such that the overall phase difference $\Delta\Phi = \varphi_{p2} - \varphi_{p1} + \Delta\theta$ can be considered as an uncorrelated random variable. Moreover, we will assume further that $\varphi_{p1}$ and $\varphi_{p2}$ (and with them $\Delta\Phi$) exhibit normal distribution with the rms $\sigma_\varphi = 2\pi$ [9].

In a gain-switched laser, $\Delta t$ introduced in Eq. (3) exhibits fluctuations, which are usually referred to as a time jitter. The main source of the jitter here are fluctuations of the pump current pulse front (the intrinsic jitter of pump current pulses) and fluctuations of its amplitude $I_p$ (the peak-to-peak value of the current modulation). (Under fluctuations, we understand here random variations of $\Delta t$ and $I_p$ from pulse to pulse.) The relation between the jitter and fluctuations of $I_p$ is defined by fluctuations of the delay occurring between the application of the current pulse and the emission of light (the so-called turn-on delay [18], [19]). However, at high modulation frequencies (more than 1 GHz) the carrier number $N$ does not have time to get well below threshold; therefore, fluctuations of $I_p$ cannot provide significant fluctuations of the turn-on delay and thus does not contribute significantly to the time jitter. Therefore, the main contribution to the jitter at high modulation frequencies is given by the intrinsic jitter of pump current pulses. We will assume below that fluctuations of $\Delta t$ due to jitter exhibit normal PDF; the rms of the jitter we will denote by $\sigma_{\Delta t}$.

Due to the relationship between the injection current and the shape of the optical signal, it is obvious that fluctuations of $I_p$ will lead also to random changes in the output optical power $P(t)$. If the pump current fluctuations are relatively small, one can neglect the change in the pulse shape and assume that only the "area" under the pulse varies from pulse to pulse. This fact is taken into account in Eq. (5) by parameters $p_1$ and $p_2$, each of which is a random variable with the mean value equal to $\bar{p} = 1$ and with the PDF $f_p$ defining the relationship between fluctuations of the injection current and $P(t)$. We assume that $f_p$ has the form of a normal distribution with the rms of $\sigma_p$. It is important to note that although introduced random variables $p_1$ and $p_2$ have the same mean value and exhibit equivalent PDFs with the same rms value, they cannot be substituted by a single random variable $p$, since fluctuations of $P_1(t)$ and $P_2(t)$ are independent.

We now begin to consider the PDF of the random interference signal. In order to simplify further analysis, it makes sense to get rid of the time dependence in $S(t)$ considering instead the integral signal:

$$\tilde{S} = \frac{\int_{-T/2}^{T/2} S(t) dt}{\int_{-T/2}^{T/2} P_1(t) dt}, \qquad (6)$$

where $T$ is a time window cutting out a separate pulse from the pulse train ($T$ corresponds here to the pulse repetition period), and normalization is performed with respect to the pulse exiting from the short arm of the interferometer. A further problem is then reduced to finding the PDF of the integral signal $\tilde{S}$, which we will denote by $f_{\tilde{S}}$.

In concern with integration of the signal according to Eq. (6) it should be noted that such an approach may seem not similar to how such measurements are usually performed with fast detectors and oscilloscopes. In fact, a fast digital oscilloscope with sufficiently high bandwidth (say, more than 30 GHz) will allow getting the result of the interference of chirped laser pulses even if they are shifted (i.e. delayed) relative to each other. So, it seems that if the detection is accomplished by sampling the interference signal within the pulse, the chirp will not affect the statistics of the recorded signal. However, it is not true for the case, when laser pulses are subject to significant jitter. In this case, the profile of the resulting pulse will be different for different pairs of interfering pulses, such that the pulse sampling in a certain point will anyway provide a range of values even if the phase difference between the pulses is always the same. This means that accumulating sampling points to measure PDF we will perform some kind of integration. Generally, selecting different points within the pulse to measure statistics of the interference signal, we could get various appearances of the PDF; therefore, to avoid ambiguity it seems to be more preferable to perform measurements integrating the whole pulse instead of sampling a single point. Therefore, the model based on the use of Eq. (6) is reasonable when one considers the interference of chirped laser pulses affected by jitter.

Finally, note that in a real experiment, the PDF of the interference signal is additionally "broadened" due to noises in

the photodetector. The experimental signal should be therefore written as follows:

$$\tilde{S} \to \tilde{S} + \zeta, \quad (7)$$

where $\zeta$ is the noise signal, whose probability distribution is generally considered to be Gaussian.

We found the PDF of the integral signal with Monte-Carlo simulations using the following procedure. We set the time dependence of the pump current in the form of a train of rectangular pulses, $I(t) = I_b + I_p(t)$, where $I_b$ is the bias current (the electric pulse width we denote by $w$ – see Table I). With such $I(t)$, we solved numerically rate equations (4) and selected a pulse within the time window $[-T/2, T/2]$ (one pulse repetition period) far enough from $t = 0$, such that the selected pulse was not affected by transients. Then for each set of random values $\Delta t$, $\varphi_{p1}$, $\varphi_{p2}$, $p_1$, and $p_2$ we calculated $S(t)$ according to Eq. (5), assuming that $T_{01}T_{10} = T_{02}T_{20} = 0.25$ (an ideal coupler) and $a_1 = a_2 = 0$. (Note that to fit experimental data we use below $a_1 \neq a_2$.) At each iteration, we calculated the value of $\tilde{S}$ according to Eq. (6). $10^5$ iteration were found to be enough to get quite detailed statistics. Common laser and pump current parameters used for simulations are listed in Table I; the rest parameters varied depending on the simulation.

We will now consider three different models (see Fig. 2) to show the contribution of various effects. The first model (Fig. 2(a)) corresponds to the case of $I_b = 7$ mA, $I_p = 10$ mA, $\chi = 25$ W$^{-1}$, and $\alpha = 0$ (other parameters were taken from Table I). The photodetector noise was included according to Eq. (7) with rms $\sigma_\zeta = 0.05$. One can see that the laser pulse in this case has a bell-type shape, which, with a good accuracy, can be represented by the Gaussian function. Since the linewidth enhancement factor was put to zero, the rate equation for the phase yields: $d\varphi/dt \equiv \Delta\omega = \omega - \omega_0 = 0$, i.e. the laser pulse is chirpless. One can easily find the integral signal $\tilde{S}$ of the Gaussian chirpless pulse according to Eqs. (5) and (6):

$$\tilde{S} = s_1 + s_2 + 2\eta_{\Delta t}\sqrt{s_1 s_2}\cos\Delta\Phi, \quad (8)$$

where $s_1$ and $s_2$ are normalized integral signals exiting from the short and long arms of the interferometer, respectively, and the visibility $\eta_{\Delta t}$ is given by $\eta_{\Delta t} = \exp(-\Delta t^2/8\delta^2)$, where $\delta$ is the rms width of the laser pulse. Normalized signals $s_i$ are related to random variables $p_i$ introduced above in the following way: $s_1 = p_1$ and $s_2 = rp_2$, where

$$r = \frac{(1-a_2)T_{02}T_{20}}{(1-a_1)T_{01}T_{10}}, \quad (9)$$

According to the above assumption ($a_1 = a_2$), we have $r = 1$, therefore, the rms of fluctuations of $s_i$ and $p_i$ can be assumed to be the same, i.e. $\sigma_s = \sigma_p$. One can see from Fig. 2(a) that the PDF exhibits noticeable asymmetry: the left maximum is higher and "thinner" than the right one. This feature is due to fluctuations of normalized amplitudes $s_1$ and $s_2$ and it becomes more pronounced when increasing the rms value of these fluctuations.

TABLE I
LASER AND PUMP CURRENT PARAMETERS, USED FOR SIMULATIONS IN FIGS. 2 AND 3. THE REST PARAMETERS VARIED DEPENDING ON THE SIMULATION.

| Laser | Value | Pump current | Value |
|---|---|---|---|
| $N_{th}$ | $6.5 \times 10^7$ | $w$ | 200 ps |
| $N_0$ | $5.0 \times 10^7$ | $\omega_p/2\pi$ | 2.5 GHz |
| $C_{sp}$ | $10^{-5}$ | $\sigma_{\Delta t}$ | 10 ps |
| $\Gamma$ | 0.12 | $\sigma_p$ | 0.05 |
| $\varepsilon$ | 0.3 | | |
| $\tau_e$ | 1.0 ns | | |
| $\tau_{ph}$ | 1.0 ps | | |

The picture is more complicated, if the Henry factor is non-zero. In this case, the time dependence of $\Delta\omega$ (the chirp) follows the time evolution of the carrier number $N(t)$ (see Eq. (4)), and $\Delta\omega = 0$ at $N = N_{th}$. In Fig. 2(b) we put $\alpha = 6$ without changing the bias current. One can see from the figure that $\Delta\omega(t)$ is approximately linear along the laser pulse profile (of course, it is not linear, when the contribution of spontaneous emission is not negligibly small). In fact, it is easy to find from Eq. (4) that the chirp of the Gaussian laser pulse is $\Delta\omega(t) = -\beta t$, if the spontaneous emission and the gain saturation are neglected (here, $\beta = \alpha/2\delta^2$). (If, however, $\chi \neq 0$, the time dependence of $\Delta\omega$ deviates from the straight line.) The result of the interference differs now from the chirpless case; this is clearly manifested in a change of the PDF appearance shown in Fig. 2(b), where one can see the high peak in the center. This peak indicates an increase in the probability that the signal equals $\tilde{S} = s_1 + s_2$, which is the evidence of interference worsening.

It should be noted that the difference between (a) and (b) cases in Fig. 2 is quantitative rather than qualitative in nature. In fact, the PDF of the interference signal given by Eq. (8) will also have a pronounced maximum at $\tilde{S} = s_1 + s_2$, if the rms of the jitter is quite large. So, the inclusion of the chirp increases the influence of the jitter. One can easily see this using Eq. (4) for the linearly chirped Gaussian laser pulse. For such a pulse, the integral signal $\tilde{S}$ is again defined by Eq. (8), but the visibility is given now by $\eta_{\Delta t} = \exp[-(1+\alpha^2)\Delta t^2/8\delta^2]$, which indicates the increase of the jitter by a factor $\sqrt{1+\alpha^2}$.

It is important to note that in terms of visibility the interference does not deteriorate when adding chirp to the model. Indeed, although $\Delta t$ and $\Delta\Phi$ in Eq. (8) have different values for different pairs of interfering pulses, there is a fairly high probability that the instant value of $\Delta t$ will be zero and simultaneously $\Delta\Phi = \pi$ (which provides perfect destructive interference) or $\Delta\Phi = 0$ (which provides perfect constructive interference). However, the joint probability of these events decreases when increasing jitter, which leads to an increase in the central peak in the PDF of the integral signal $\tilde{S}$. Therefore, speaking about the deterioration of interference, we do not mean a decrease in visibility but the deviation of the signal PDF from that shown in fig. 2(a).

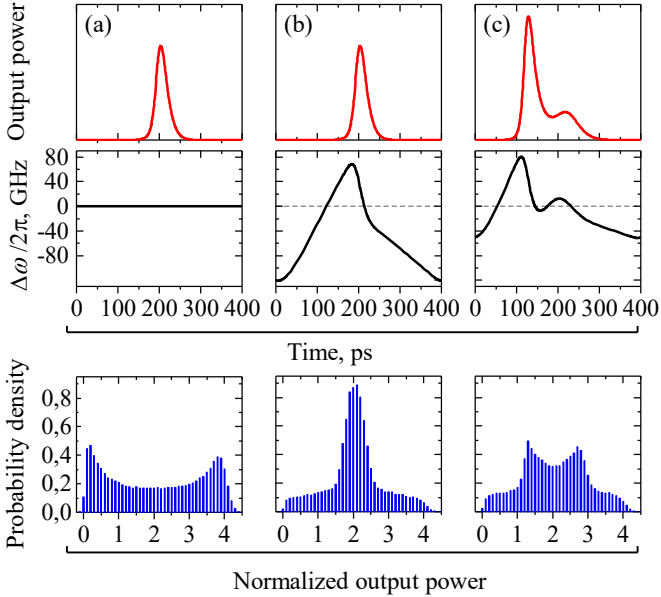

Fig. 2. The shape of the laser pulse (top), its chirp $\Delta\omega(t)$ (middle) and the PDF of the normalized interference signal (bottom) in different models: (a) chirpless ($\alpha = 0$) bell-shaped laser pulse; (b) chirped ($\alpha = 6$) bell-shaped laser pulse; (c) chirped ($\alpha = 6$) laser pulse affected by relaxation spike. Simulation parameters are listed in Table I.

Finally, in Fig. 2(c) we consider a more general case, when the optical pulse is distorted by the first relaxation spike. For this, we increased the value of the bias current up to $I_b = 9$ mA ($\alpha$ was the same as in the previous simulation in Fig. 2(b)). One can see that due to the asymmetry of the output power $P(t)$, the chirp $\Delta\omega(t)$ has a quite complicated form. The PDF of the interference signal in Fig. 2(c) exhibits two pronounced maxima, which, in contrast to Fig. 2(a), do not correspond to an ideal constructive and destructive interference.

## III. EXPERIMENT

We will now proceed to laser pulse interference measurements demonstrating obtained theoretical results. The optical scheme used in this work to observe interference of laser pulses is shown in Fig. 1. Unbalanced fiber optic Michelson interferometer was built using an optical circulator, a 50:50 single mode (SM) fiber coupler, SM fiber patch cable as a delay line, and two Faraday mirrors used to compensate the effects of polarization mode dispersion in SM fiber components. The length of the delay line $\Delta L$ was calculated using the following formula:

$$2\Delta L = \frac{2\pi N_p c}{\omega_p n_g}, \qquad (10)$$

where $c$ is the speed of light in vacuum, $n_g$ is the group index, $\omega_p$ is the current modulation (angular) frequency corresponding to the pulse repetition rate, and $N_p$ is the number of pulses emitted by the laser during the time when the given pulse travels the distance $2\Delta L$. In our case, $\Delta L$ was 128 cm, which at $\omega_p/2\pi = 2.5$ GHz provides $N_p = 32$, such that the first laser pulse interferes with the 33$^{rd}$ one, the second pulse interferes with the 34$^{th}$ one, etc.

The 1550 nm telecom distributed feedback (DFB) laser with 10 Gbps modulation bandwidth was driven by a commercial 11.3 Gbps low-power laser diode driver. Thermal stabilization of the laser diode was performed using Peltier thermoelectric cooler controlled by commercially available single-chip temperature controller. The waveform modulated at 2.5 GHz was generated by a phase-locked loops multiplying the input frequency from the 10 MHz reference oscillator. The peak-to-peak value of the modulation current $I_p$ was estimated to be ~10-12 mA. The laser threshold current $I_{th}$ found from the light-current characteristics was estimated to be around 10 mA.

To detect the optical output, we used the home-built photodetector equipped by a p-i-n photodiode with 10 GHz bandwidth. The signal processing was performed using the Teledyne Lecroy digital oscilloscope (WaveMaster 808Zi-A) with 8 GHz bandwidth and temporal resolution of 25 ps. Optical spectra were acquired using Thorlabs optical spectrum analyzer (OSA 202) with a spectral resolution of 7.5 GHz.

Experimental PDFs of the interference signal at four different values of the bias current $I_b$ are shown in Fig. 3 by red circles. Corresponding simulations are shown by blue histograms. For simulations, we used parameters from Table I. The rms of the normalized detector noise we put to $\sigma_\zeta = 0.25$, the gain compression factor was $\chi = 30$ W$^{-1}$, and the peak-to-peak value of the pump current was $I_p = 11$ mA. As above, initial phases of laser pulses, $\varphi_{p1}$ and $\varphi_{p2}$, were assumed to exhibit normal distribution with $\sigma_\varphi = 2\pi$. Finally, the interferometer arms were assumed to exhibit different losses (we put $a_1 = 0$ and $a_2 = 0.1$; these values were estimated experimentally). According to theoretical consideration, the contribution of relaxation oscillations at $I_b = 6$ mA is quite small; therefore, the corresponding PDF is similar to that shown in Fig. 2(b). The PDF at $I_b = 9$ mA is substantially different from that obtained at $I_b = 6$ mA due to the higher impact of relaxation oscillations. Intermediate PDFs in Fig. 3 are presented to demonstrate its evolution from lower to higher values of the bias current.

Obviously, the "chirp + jitter" effect can be reduced by either reducing jitter or chirp, or both. In our opinion, the simplest (and cheapest) solution is to use the bandpass filter to cut off a part of the laser spectrum associated with chirp. In the case of the Gaussian laser pulse with linear chirp, this approach would be more difficult, since it is necessary to cut off both the high- and low-frequency components of the spectrum. For the laser pulse affected by relaxation oscillations, the optical spectrum will have essential asymmetry, since only the rising edge of the laser pulse will be chirped significantly. One can see from Fig. 2(c) that when the relaxation spike occurs, the absolute value of $\Delta\omega$ decreases, which makes the falling edge of the pulse less chirped. Therefore, it is enough just to cut off the high-frequency part of the spectrum in this case.

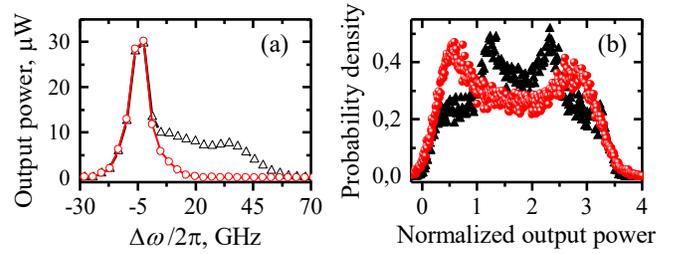

Fig. 4. (a) Experimental optical spectra at $I_b = 9$ mA without (empty triangles) and with (empty circles) DWDM filter. (b) Experimental PDFs of the interference signal at $I_b = 9$ mA without (filled triangles) and with (filled circles) DWDM filter.

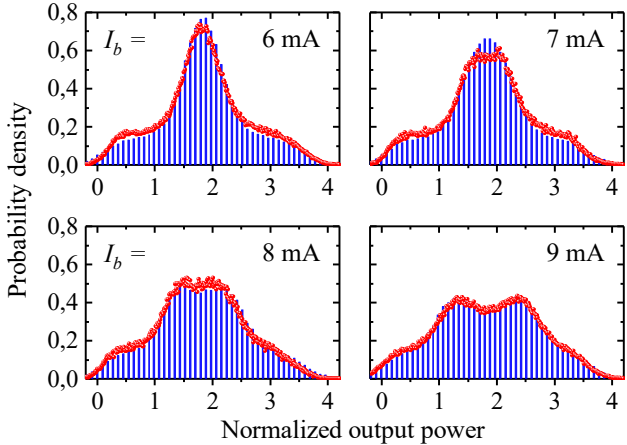

Fig. 3. Experimental PDFs of the interference signal at three different values of the bias current $I_b$ (red circles) and corresponding Monte-Carlo simulations (histograms). The peak-to-peak value of the modulation current was ~10 mA in all cases.

To cut off the laser spectrum we used the telecom dense wavelength division multiplexing (DWDM) filter with 100 GHz bandwidth placed just after the laser output (see Fig. 1). The position of the laser spectrum on the frequency axis was adjusted by changing the laser temperature in such a way that the high-frequency shoulder was beyond the filter bandwidth. Experimental optical spectra at $I_b = 9$ mA without and with DWDM filter are shown in Fig. 4(a) by empty triangles and empty circles, respectively. The central frequency $\omega_0$ corresponds to the "center of gravity" of the filtered spectrum at given temperature and is $\omega_0/2\pi = 193.63$ THz. One can see that the unfiltered spectrum has a broad high-frequency shoulder, which is related to the laser pulse chirp. Indeed, the corresponding PDF shown in Fig. 4(b) by filled triangles exhibits the specific shape caused by the interference of chirped non-Gaussian laser pulses (Fig. 2(c)). The PDF obtained with the DWDM filter (filled circles in Fig. 4(b)) exhibits two pronounced maxima corresponding to the constructive and destructive interference, as for the model shown in Fig. 2(a).

It should be noted here that the spectral filtering does not change the chirp itself – it only changes the intensity distribution of spectral components in the pulse. In fact, we observed a decrease in the intensity of the rising edge of the laser pulse after passing the optical filter, which is caused by the fact that its rising edge is chirped more significantly than the falling one. So, an optical filter improves spectral matching of the pulses improving thus their interference.

## IV. CONCLUSIONS

We developed a detailed model for the interference of optical pulses from a gain-switched semiconductor laser in the presence of chirp, jitter, and relaxation oscillations. The model allows explaining evolution of the signal PDF against the pump current, which will be helpful for analysis of QKD systems and optical QRNGs. We demonstrated that chirp, jitter and relaxation oscillations have a significant impact on probabilistic properties of the interference of laser pulses. It was shown that the relaxation spike makes the falling edge of the laser pulse less chirped and thus reduces the impact of the "chirp + jitter" effect on the appearance of the signal PDF. Moreover, the optical spectrum of the chirped pulse accompanied by relaxation oscillations exhibits significant asymmetry and can be easily cut off with a bandpass filter.

Note that in the context of a QRNG, the jitter should be considered as a source of a "classical" noise, since it is mainly caused by fluctuations of the pump current. Quantum noise originating in spontaneous emission and amplified via the pulse interference is thus "contaminated" by the jitter. Therefore, the combined effect of the chirp and jitter is crucial when elaborating QRNG and must be minimized, e.g. with the use of the spectral filtering.

## ACKNOWLEDGMENT

We are grateful to Aleksey Fedorov and Denis Sych for valuable comments.